\documentstyle{aipproc2}
\begin{document}
\title{ Cosmological Gravitational Wave Backgrounds}

\author{Craig J. Hogan}
\address{Astronomy and Physics Departments,
University of Washington
\thanks{This work was supported at the University of
Washington by NASA.}, Seattle, WA 98195-1580 }

\maketitle

\begin{abstract}
 An overview is presented of possible cosmologically distant
sources of gravitational wave backgrounds, especially those which might
produce detectable backgrounds in the LISA band between
0.1 and 100 mHz.
Examples considered here include
inflation-amplified vacuum fluctuations in inflaton  and
graviton fields, bubble collisions in first-order phase 
transitions,  Goldstone modes of classical self-ordering
scalars, and cosmic strings and other gauge defects. Characteristic scales and
 basic mechanisms are
reviewed and spectra are estimated for each of these
sources. The unique impact of a LISA detection on fundamental physics
and cosmology is discussed.
\end{abstract}

\section{Introduction}

In relativistic Big-Bang cosmology, the Universe is optically
thin to gravitational waves all the way back to the Planck epoch;
once a wave is produced, absorption is negligible, with energy
losses due to redshift alone.  The
cosmological background of gravitational waves thus contains directly
observable information from the entire history of the
macroscopic  universe--- an unobstructed view of  our  past
light cone.
Here I present a broad survey of
 ideas for processes in the early universe
which may produce   gravitational waves, including guesses at their
amplitudes and frequency spectra, and 
what a  detection by LISA might teach us about
the early universe.  I direct the reader to other recent
documents and reviews for more detailed and
comprehensive descriptions of many of these ideas and for
more thorough surveys of the literature.    
\cite{gw96,gw96b,gw98,lisa98}

  As in so many areas of astrophysics, gravitational waves offer
insights into cosmological processes quite distinct from
those accessible by electromagnetic observations. Direct light
paths come to us from a temperature $T\approx 3000K$ and a cosmic age  
$t\approx 0.5My$; the thermal spectrum of the primordial radiation was
fixed   when the last efficient photon-production processes froze
out at   $T\approx 500 eV$ and a cosmic age of a few weeks. 
By contrast, the LISA gravitational wave backgrounds, if they
exist, illuminate much earlier events, and probe directly the physics of very 
high energies: the electroweak, GUT, and Planck scales.
LISA is sensitive to 
gravitational waves produced by cosmic strings in the temperature
interval 10 keV--- 10MeV, and cosmic ages from $10^{-2}$ to $10^4$ 
seconds; our current knowledge of these epochs comes only from fossil
low-bandwidth information such as the light element abundances.
 Large-scale
relativistic flows of energy during
phase transitions above 100 GeV or so (earlier than
about $10^{-10}$ seconds) produce waves in the
LISA band but have few other observable effects; it is possible that  
production of these waves may be associated with our only other relics of these
epochs, the cosmic baryon number and perhaps the dark matter. The 
physics which shaped the metric on the largest scales of space
and time, currently
described by inflation models, probably also generated gravitational waves
in ways akin to the generation of the perturbations which led to cosmic
structure, perhaps at the GUT epoch. There are also
ways in which   many spacetime dimensions and
other internal degrees of freedom now being 
explored for fundamental physics near the Planck
scale could lead to intense gravitational wave backgrounds in 
the LISA band, produced  at temperatures of 1 to 1000 TeV. 
  Detection of a cosmological background by LISA would
provide our first view of
early mesoscopic gravitational phenomena about which no other
trace survives, with a likely connection to the frontier
problems of cosmological theory---
 the production
of  baryons,   dark matter, and 
fluctuations in binding energy which led to cosmic structure---  
as well as the structure of   fundamental fields at 
very high energies.

\section{Redshifted Hubble Frequency}

As a reference point for the following discussion, it is useful
to define a characteristic frequency associated with most classical
production mechanisms for cosmological  gravitational waves, the Hubble
rate divided by the cosmological redshift:
\begin{equation}
 \omega_0(z)\equiv {H(z)\over (1+z)}\approx 2\times 10^{-5}{\rm Hz}
{T(z)\over 100{\rm GeV}},
\end{equation}
 where $T$ is the temperature of the universe.
(There is a weak dependence $\omega_0\propto g_*^{1/4}$
on the number of effective
degrees of freedom $g_*$ which   varies only slowly in 
the Standard Model, 
 from about $g_*\approx 60$ at 1 GeV to about 100 at 1 TeV.)
This is the frequency of gravitational waves observed today 
which were produced on the horizon scale at temperature $T$.
 Gravitational
waves in the LISA band can be produced by horizon-scale processes at 
temperatures between a few hundred GeV and a few hundred TeV,  by
relativistic processes well within the horizon at lower temperature,
or by inflationary processes at much higher temperature.

\section{Broad-Band  Energy}

We specify the gravitational wave spectrum $\Omega_{GW}(f)$ in terms of
energy  density per
unit of log frequency, in units of the critical density.
After gravitational waves are produced they redshift in the same way
as electromagnetic waves and other relativistic forms of energy. Therefore,
the integrated energy density of the gravitational wave background(s) scale in proportion
to the energy density of the sum of the other relativistic 
components. Currently this sum is (assuming massless neutrinos)
\begin{equation}
\Omega_{rel}(z=0)=
\Omega_{\gamma+\nu+\bar\nu}
=0.7\times 10^{-4} T_{2.726K}^4 h_{75}^{-2}.
\end{equation}
We define a  quantity 
\begin{equation}
F(f)=\Omega_{GW}((1+z)f)/\Omega_{rel}(z),
\end{equation}
the ratio of gravitational wave to other relativistic energy,
which is approximately conserved. Notice that at high redshift
$\Omega_{rel}$ is shared among many more relativistic degrees of
freedom than it is today, but $F$ is conserved so long as the
coupled system of particles remains highly relativistic.

The quantity $F$ roughly corresponds to the overall gravitational wave
production efficiency--- the fraction of mass-energy converted
into gravitational waves.
Since it seems unlikely that gravitational waves would be produced
 more efficiently than other forms of energy, a plausible upper limit
on the cosmological background is $F<1$ at all frequencies. 

Compare this with the LISA sensitivity as estimated in
the Yellow Book\cite{lisa98}.  The rms amplitude of a fluctuating gravitational
wave in a bandwidth $f$ about a frequency $f$ is
\begin{equation}
h_{rms}(f, \Delta f=f)= 10^{-15} [\Omega_{GW}]^{1/2}
f_{mHz}^{-1}h_{75};
\end{equation}
however   the background will be distinguishable 
from instrumental noise only over much narrower bands; the strain
produced in one frequency resolution element after a year
of observation is
\begin{eqnarray} 
h_{rms}(f,\Delta f=3\times 10^{-8}{\rm Hz})
&=&5.5\times 10^{-22} [\Omega_{GW}/10^{-8}]^{1/2}f_{mHz}^{-3/2}h \\
& & =4.6\times 10^{-20} F(f)^{1/2}f_{mHz}^{-3/2}h~.
\end{eqnarray}
At a $f\approx$ few mHz, LISA's instrumental noise drops to as low as
$h_{rms}\approx 10^{-24}$ in this band so one can contemplate detecting
effects as small as 
\begin{equation}
F\approx 10^{-8}$ ~~{\rm or} ~~ $\Omega_{GW}\approx 10^{-12};
\end{equation}
indeed for backgrounds larger than this the cosmological background starts
to dominate the noise over some frequency intervals, becoming a nuisance
for other observations.

\section{Current Observational Constraints on $\Omega_{GW}$}

Although within its band LISA achieves
a sensitivity in $F$ and $\Omega_{GW}$ far 
better than any other technique, we  already have some meaningful constraints
on $\Omega_{GW}(f)$ in other frequency intervals which impact 
the candidate sources for LISA.

The most sensitive in terms of $F$ is cosmic background radiation
 anisotropy.\cite{cbr,cbr2} Tensor mode perturbations generate temperature fluctuations
in the microwave background; on scales larger than about a degree
these preserve  roughly
the amplitude they had entering the horizon
$\delta T/T\approx h_{hor}$.  Now $\Omega_{GW}\approx (f/H_0)^2h_{now}^2$,
so
observed limits (and measurements, which may not be of 
tensor modes) of
about
$\delta T/T\approx 10^{-5}$ yield a constraint $\Omega_{GW}(f=H_0)
\lesssim
10^{-10}
$ on the current horizon scale. The limit becomes
 smaller at higher frequencies,
the details depending on the cosmological model. Detailed constraints
are placed on inflation models from the predicted tensor modes and
their effect on anisotropy.
In principle, though not yet in practice, polarization allows
tensor and scalar sources to be distinguished observationally.
\cite{polardefect97,polar98}

 A scale-free 
process such as inflation with $h_{hor}=$ constant creates
a  flat background spectrum
above about
$f=(\Omega_{rel}(z=0)/\Omega)^{1/2}$,
with $F(f)$= constant $\approx h_{hor}^2\approx 10^{-10}$
with amplitude constrained by CBR anisotropy.
In the scale-free case,  CBR data  
 provide  the most sensitive limits on $F$--- better than LISA.
 However, it is still
interesting to consider direct limits on waves at higher frequencies
since   sources are never precisely  scale-free and sometimes
not even approximately so.
 
Pulsar timing measurements directly limit the background at
frequencies determined by  the observation timescale of a 
few years. The principle resembles that of interferometric
detectors.  The pulsar  acts as a very steady  clock with 
timing residuals $\delta t$ of about a microsecond, and over
few years the lack of deviations from 
steady ticking (aside from those expected from 
Newtonian accelerations of both us and the pulsar)
 constrains the strain amplitude to
$h\lesssim \mu sec/ 10^8 sec\approx 10^{-14}$. After allowing for
the fact that unknowns such as the
 Newtonian accelerations and the precise pulsar direction
are ``fitted out'', the current limit\cite{pulsar94} is  
 $\Omega_{GW}(f=10^{-8} {\rm Hz})h^2 \le 6\times 10^{-8}$. Note that
  if many   pulsars
are added with accurate timing and good coverage over the sky,
it is possible to extract a signature indicating a positive
detection of gravitational waves.

For backgrounds that were already present at the epoch of cosmic
nucleosynthesis $1 sec<t<100 sec$, abundances of light elements
provide another constraint. The presence of gravitational waves
adds to the total energy density of in the same way as adding additional
relativistic degrees of freedom; for example, an extra neutrino
species adds the equivalent of $F\approx 1/6$.  Although the precise
limits are a matter of opinion riding upon continually changing debate
over observational errors
\cite{sbbn95,sbbn96,sarkar96,sbbn97,sbbn98}, it is clear that standard nucleosynthesis
fails (primarily due to overproduction of helium-4) unless
$F\lesssim 0.1$.  This limit applies to most of the sources we will consider 
for the LISA band. Notice that the limit becomes stronger for
scale-free backgrounds for which many octaves of $f$ contribute
to the density.

Chaotic universes with $F\approx 1$ are prone to forming many
black holes. This is a disaster since the energy locked up
in black holes does not redshift away and   quickly comes
to dominate the energy density.\cite{bdm94}  At most $10^{-8}(T/GeV)^{-1}$
of the mass can convert can convert to holes
at temperature $T$ without exceeding the mass per photon
in the universe today. So there is constraint on the production
mechanism for backgrounds which approach the $F\approx 0.1$
nucleosynthesis bound--- they must produce their waves efficiently
in a well-regulated process that does not allow a wide dispersion
of gravitational potentials since even a tiny fraction of matter
in very deep potentials ($v\approx 1$) causes problems.
 The Goldstone modes discussed
below offer an example of such a  mechanism but the phase transition
bubble collisions do  not.

There are also limits from the spectrum of the background
radiation. Gravitational waves 
  create observable quadrupoles in the radiation field at each point, so
 the average radiation field is no longer thermal but a mixture
 of temperatures with a spread of the order of $h_{rms}$ 
times the Hubble velocity $H(z)/f$
for $f$ comparable to the scattering rate of photons
at redshift $z$.
The COBE/FIRAS limits on spectral distortions limit energy
inputs after this epoch to the order of
$10^{-4}$\cite{cbr2}; but the corresponding limit on $h_{hor}$  is competitive with 
the CBR anisotropy limits only over a narrow range of frequencies.

\section{Gravitational Waves from Inflation}

Inflation generates gravitational wave backgrounds by the 
parametric amplification of quantum fluctuations, the same 
process thought to create the scalar modes that lead to
large-scale cosmic structure.\cite{inflation97,inflation97b}
Backgrounds are created both by the fluctuations of the inflaton
field--- which make  the familiar scalar modes---
and by quantum fluctuations of the gravitational field itself.

The amplitude of scalar perturbations depends on
details of the  inflaton potential. These are  tuned to yield $h_{scalar}$
in agreement  with some suite of observations including
the CBR fluctuations and large scale structure.
The best fit to the data 
for scalar modes  is  close to scale-free with amplitude
$h_{hor,scalar}\approx 10^{-5}$, but may have a small
``tilt'' or slow variation of $h_{hor}$ with scale.  As they 
enter the horizon, there is some mixing between
scalar and tensor modes, since the scalar perturbations
lead to quadrupolar mass flows that act as sources for
gravitational waves. Very roughly,
the mixing will lead to  $h_{hor,tensor}\approx h_{hor,scalar}^2$.
In spite of the suppresion, the LISA band is so much higher
in frequency 
than the direct observational constraints that even a small  tilt 
can lead to essentially any amplitude in the LISA band.

 In the case
of gravitational field fluctuations, the amplitude is not
dependent on the inflaton potential directly but essentially
on just the expansion rate or density during inflation
when waves of observed frequency $f$ match the inflationary
expansion rate:
\begin{equation}
F(f)\approx V_{inflation}(f)/  m_{Planck}^4
\end{equation}
which is close to scale-free and hence limited to
$F\lesssim 10^{-10}$. Inflation very close to the Planck
scale generally runs into difficulties with CBR anisotropy from
these tensor sources.

The conclusion is that LISA may detect gravitational waves
from inflationary fluctuations (from the inflaton fluctuations)
if the tilt of the spectrum is favorable, but
is unlikely to detect modes directly generated by quantum fluctuations
of the graviton.

\section{First-order Phase Transitions:
Relativistic Fluid Flows and Bubble Collisions}

The universe may have undergone catastrophic phase transitions
at various stages, associated with a sudden change in the
ground state configuration of the vacuum fields.\cite{peierls52,nucleation83,transition86,transition90,transition93,transition94,transition96}
The macroscopic
 description is similar whether the fields are associated with
 QCD, electroweak breaking, or supersymmetry breaking. 

    We imagine an
order parameter 
$\phi$ with a free energy density or effective potential $V_T(\phi)$;
this potential has two distinct minima corresponding to
two distinct phases; the free energy difference between them
vanishes at the critical temperature $T_c$, one phase being favored at
higher and the other at lower temperature. The transition
from one phase to the other cannot happen smoothly because
of the activation barrier between them corresponding to
the energy cost of creating a surface interface between phases. The system
supercools by a small amount  until the  free energy  difference
between phases is sufficient to nucleate bubbles; thermal or quantum
fluctuations must create a bubble of low-temperature phase large enough
that the volume energy difference exceeds the cost of creating
the surface
between them; above this size bubbles grow by detonation or 
deflagration with the phase boundary propagating close to 
the speed of light. 
The release of latent heat heats the inter-bubble medium back to almost
$T_c$ and increases the pressure
between bubbles, so (in the deflagration case) after the shocks from 
the bubbles meet, fresh nucleation slows and the transition
finishes by the slow growth of  already-nucleated  bubbles. The universe
can expand  for a significant quiescent
 period near $T_c$ with both phases coexisting;
as it expands it fills more with the lower-density, low-temperature
phase. The remnants of the high phase are eventually isolated
as islands by the percolation of the low phase,
and finally when  there is no more high phase left normal
cooling resumes.

The production of gravitational waves occurs because of relativistic
flows of matter of different densities in the two phases; a substantial
fraction of the matter is accelerated to close to the speed of light, and
asymmetric shocks are formed as bubbles collide. Not all of the processes
involved are computed accurately and many depend on the detailed input
physics, but the main parameter is generic: the maximum fractional supercooling
$\delta$. Since this is the amount by which the universe expands
during the nucleation of the typical bubbles, it determines the bubble
size $\delta/H$. Because the nucleation rate depends exponentially
on $\delta$, even a very strongly first order transition generically
obeys\cite{nucleation83}
\begin{equation}
\delta\lesssim \log [T/m_{Planck}]\approx 10^{-2}
\end{equation}

From scaling we estimate the background from flows of  
scale $\delta/H$, maximal density contrast, 
 and $v\approx c$; it will be a broad-band
background of with a peak at frequency $f_{peak}\approx \omega_0(T)/\delta$
and  peak amplitude $F(f_{peak})\approx \delta^2$.

For example, if there is a very strongly first-order phase
transition at 100GeV to 1TeV, a background could be generated with
a characteristic frequency of around 2 to 20 mHz and an amplitude
as large as $\Omega_{GW}\approx 10^{-8}$, which is detectable. These
parameters might be associated with electroweak symmetry breaking and/or 
supersymmetry breaking. Although a first order transition is 
not required by the Standard Model, many workers believe that
it is first order because that or some other significant
disequilibrium is required to create the baryon asymmetry. There
is at least the possibility that LISA might make an important
connection here, the first direct window on the process that
created cosmic matter from radiation.

On the other hand, a very strong disequilbrium may not be required
and the fractional supercooling   could easily be orders of magnitude
less than $\delta\approx 10^{-2}$, which would make the background undetectable.
This   seems  likely to be the case for the QCD transition
(which is probably not even first order,
but had it been strong
 might have been detected at lower frequencies from pulsar timing).

\section{Self-Ordering Scalars: Gravitational Waves from
Goldstone Modes}

Gravitational  waves which may be generated by global excitations of 
new classical scalar degrees of freedom. Such fields often
appear in effective theories derived from unified models such
as supersymmetric theories and string theores.

We   describe the behavior of 
active classical scalar fields
with  the simple Lagrangian density
\begin{equation}
L= \partial_\mu\phi\partial^\mu \phi/2-V(\phi)
= \dot\phi^2/2-V(\phi),
\end{equation}
leading to the
evolution equation
\begin{equation}
\ddot \phi+3H\dot\phi-\nabla^2\phi+{\partial V/\partial \phi}
=0,
\end{equation}
 We now suppose that there is more than one scalar component
and that the
effective potential $V(\vec\phi)$ has some set of degenerate minima,
  no longer at just one
 $\phi $  but over some set of points far from the origin
which all have $|\vec\phi|=\phi_0$.
For each direction within this surface with ${\partial V/\partial \phi}=0$
this wave equation describes massless ``Goldstone modes'',
coherent classical massless modes which propagate at $c$ and
dissipate only by redshifting (via $3H\dot\phi$).\cite{gold82,gold95}

 Typically different states
$\vec\phi$ are
reached at different points in space
 by cooling down from some higher symmetry ({\it a la} Kibble)
which generates spatial
gradients in 
   $\vec\phi$.
 These
variations excite the  Goldstone modes  of the field,
in general   with a  large initial amplitude, $\delta\phi\approx \phi$,
and with random phase on all scales larger than the horizon.
 Modes larger than
the horizon are essentially frozen in amplitude.  
The dominant energy density comes from modes just entering the
horizon scale (when they have propagated about one wavelength),
at which time they   contribute a 
density of the order of
$(\phi_0/m_{Planck})^2$ times the total density;
 after this time the amplitude and
frequency of the waves redshift like other relativistic waves,
and eventually dissipate.  Since the waves induce coherent
quadrupolar flows of energy on the horizon scale
and close to the speed of light, they
couple   to gravitational radiation 
 and create a gravitational wave background.
Roughly a fraction  $(\phi_0/m_{Planck})^2$ of  the scalars' 
energy is radiated per oscillation time in gravitational waves
on the horizon scale.
If the other couplings of the field are not very weak the main
energy loss may not be gravitational radiation 
and the
  gravitational wave background may be as small as
$F(f=\omega_0(z))\approx (\phi_0/m_{Planck})^4 $.
(Uncertainty arises here not only from other couplings in the 
Lagrangian but also from the gravitational coupling
to the other cosmic matter fluids, which may be dissipative and
reduce the final energy in the gravitational wave channel.)
A scale $\phi_0$ near the Planck scale is needed to produce
a detectable background.

Unless some other coupling is
added to damp the Goldstone oscillations at some point, this is a scale-
free background and is subject to the constraints discussed above
from lower frequencies, which already imply a fairly small
background.  This  constraint 
can be avoided however  if the theory contains fields which
strongly damp the Goldstone modes after a certain epoch $t$ which 
then reduces the low-frequency gravitational waves
(i.e., those below $\omega_0(t)$).  For example, a second phase transition
could occur removing the degeneracy in the minima of $V$; the fields
would everywhere relax to the single minimum, removing the source
of subsequent Goldstone excitation.

 The excitation
of these modes happens on a timescale
 determined by the motion in $\vec\phi$ space normal
to the surface of degenerate minima. If the Higgs masses
corresponding to these directions of $V$ are very small, the
Kibble excitation  may not occur until temperatures much lower than
$\phi_0$, which cuts off the spectrum at high frequencies.

Note that although the waves are generated classically at
the 1--- 1000 TeV temperatures characteristic for $\omega_0(T)$
to lie in the LISA band, the physics probed is on the scale
of $\phi_0$ which can be close to the Planck scale and reflects
new fundamental fields close to the scale of quantum gravity.
Multitudinous   internal spaces and dimensions are now being
contemplated for fundamental theory near the Planck scale
(``M-theory'' or string theory)\cite{mtheory98}. The ground state is far
from being understood but is often described using the 
kind of effective theory we have just sketched with a large
set of degenerate minima and
 many internal degrees of freedom. Since the compactifications and
symmetry breakings occur close to the Planck scale,
we might expect the effective theory to contain scales $\phi_0$
close to $m_{Planck}$. In this case the Goldstone
modes are a
plausible mechanism which may   come close to saturating the $F\approx
0.1$ (nucleosynthesis)  bound in the LISA band--
a spectacularly strong background
\begin{equation}
\Omega_{GW}\approx 10^{-5}
\end{equation} 
which could have a signal-to-noise of $10^7$!
Such a strong  detection  would clearly enable many details
to be studied and provide 
spectacular direct probe into degrees of  freedom not seen in any other
way.  Although a classical macroscopic process, it would reveal fields linked to the
unification of gravity and other forces. Even though it is
quite possible that nothing of the sort occurs near enough to
the Planck scale to produce a detectable background, we should
bear this possibility in mind.

\section{Cosmic Strings}

Strings are topologically stable defects in gauge fields, analogous
to vortex lines in superfluids, within which the vacuum is trapped
in the excited ``false vacuum'' state.   They are
formed again by the  Kibble mechanism: the rapid quenching that occurs from the
cosmic expansion prevents a global alignment of fields and guarantees
plentiful defects. After forming they stretch, move, interconnect, form
kinks and wave excitations,  and break into
loops in a complicated network teeming at close to the speed of
light. Their main energy loss is by gravitational radiation.  
\cite{strings81,strings84,strings85,strings94,strings97,strings98}

The gravitational interactions of strings are determined by
a parameter $\mu$, the mass per unit length, or equivalently
a ``deficit angle'' $\delta=4\pi G\mu$ for the conical
space created by a  straight string.
 For a symmetry breaking at scale
$m$, $\mu$ is of the order $m^2$, leading to  $\delta\approx
(m/m_{Planck})^2$. Formation of strings
is quite generic and may occur even during electroweak symmetry
breaking if the topology of the Higgs sector is suitable, but
the gravitational effects are usually only considered for GUT scale
strings with $\delta\approx 10^{-5}$    which are
heavy enough to produce large scale structure and CBR anisotropy.
Current calculations show that strings predict a poor fit to
these two datasets (too little structure for a given anisotropy
\cite{defectCMB97})
but one should bear in mind that strings may still exist
at smaller $\delta$ and produce gravitational waves. 
Strings have many very distinct observable effects;
for example, a string in the plane
of the sky creates a duplicated strip of images
of width $\delta$; galaxy images on the boundary
of the strip have sharp, straight edges.

Although
the early calculations\cite{strings84,strings85}
 of the spectrum of the background were based
on a rather simplified picture of the network, they agree 
remarkably well with recent predictions based on sophisticated
simulations of the  behavior of the network.
In the LISA band the spectrum is flat with $F\approx \delta^{1/2}$;
this is so strong that it is easily observable even if $\delta$
is too small to  affect any other astrophysical observable.
These may be the one type of object for which gravitational
radiation is the most easily observable gravitational effect!
Indeed even now the pulsar timing bound on gravitational
waves is of comparable significance to CBR fluctuations in
constraining $\mu$. 

The waves are produced by decay of string loops and kinks
which occurs after about $\delta^{-1}$ oscillations, dominated
at temperature $T$ by
  frequencies about $\delta^{-1}\omega_0(T)$; the waves in
the LISA band therefore were emitted at temperatures from 10 MeV down to
about 10 keV.

Cosmic strings and Goldstone modes both rely  on
macroscopically excited  new
scalars excited by the Kibble mechanism. However, there are
important differences.
Strings derive from  a gauge field (a local rather than a global symmetry
breaking) and a nontrivial topology in the manifold of degenerate
vacua (leading to the topological stability).
A global field with nontrivial topology is also possible 
(global strings, monopoles, textures) with qualitatively 
similar results to the Goldstone estimates.  In terms of gravitational
wave production, strings are more efficient for a given
$\phi_0$ and make cleaner predictions for a wide variety of
phenomena; on the other hand the Goldstone modes can occur
with $\phi_0$ close to the Planck scale and therefore can 
produce the most intense backgrounds.  Gravitational 
waves can also be efficiently produced by ``hybrid'' defects.\cite{hybrid96}

\section{ Impact of a Detection  on Cosmology}
All of the sources considered above are well motivated from
some physical point of view.  However,
the amplitudes for many of them are almost unconstrained.
It is possible that LISA will never detect a cosmological 
background; it is also possible that an intense cosmological 
background dominates the LISA noise budget by a large factor,
limiting its utility for studying many sources at low
redshift. In the latter case,   there will at least
be a big payoff in completely  new knowledge of the early universe. 
I have not discussed here the problems of distinguishing
backgrounds from noise or separating cosmological backgrounds
from others such as Galactic binaries. But   assuming
a cosmological background is detected, how are we to 
interpret it?

 The sources 
discussed here 
  all produce highly confused  isotropic 
stochastic backgrounds of  broad-band
Gaussian noise.  The 
 first clue to interpretation will be the shape of the
spectrum. The spectra of the sources we have considered fall into two
broad categories: 
(1) Scale-free sources with 
$\Omega_{GW}$ approximately independent of $f$, or $h_{rms}\propto f^{-3/2}
\Delta f^{1/2}$,
over the LISA band.
 These include inflation, generic Goldstone modes,
and cosmic strings. Even a tilted spectrum from inflation---
the only inflationary contribution
 likely to be detectable--- will be approximately flat
over the LISA band.  (2) Other sources with the imprint of 
some characteristic scale. These include some Goldstone models
(those where features imprinted by damping or Higgs modes 
happen to lie in the LISA band), and waves from
bubbles or other relativistic flows which will bear the imprint
of the nucleation scale where the spectrum peaks. Distinguishing
between these broad categories is possible with even moderate
signal-to-noise detection because of the fairly broad band
available, about a factor of a thousand in frequency. 
The intensity of the spectrum may give another 
clue: for example, a very intense scale-free background points to some kind of 
macroscopically active scalar.

 LISA surveys
a domain of cosmological history which has left  few other
direct observables. It is   worth commenting that these other
observables concerning the very early universe are either very large-scale
(e.g., fluctuations on
 galaxy clustering scales and above from inflation)
 or very small-scale (e.g., abundances of nuclei determined by
microscopic reaction rates and thermodynamics);
whereas gravitational waves probe a possibly   richly varied primordial
``mesoscopic'' phenomenology about which all other traces have
been erased. A detection of a gravitational wave background would 
depart from the quiescent behavior we have been led to expect from
the early universe by the observed small fluctuations, tiny
spectral distortions, and abundances in agreement with 
homogeneous nucleosynthesis; it would give us
insight into a nonlinear, chaotic or turbulent stage in  the early history
of the universe about which we currently have no clue.


\begin{references}
\bibitem{gw96}Allen, B.,
in {\it  Les Houches School on Astrophysical Sources of Gravitational Waves},
 eds. Jean-Alain Marck and Jean-Pierre Lasota,  (Cambridge University Press,
 1996); gr-qc/9604033
\bibitem{gw96b} Battye, R. A. and Shellard, E. P. S. (1996);
astro-ph/9604059
\bibitem{gw98}Flanagan, E. E., to appear in GR 15 (1999); gr-qc/9804024
\bibitem{lisa98} LISA Study Team, {\it LISA Pre-Phase A Report} (1998)
\bibitem{cbr} Bond, J. R. and Jaffe, A. H., Proc. Roy Soc. A (1998);
astro-ph/9809043
\bibitem{cbr2} Smoot, G. and Scott, D., Eur. Phys. J. 3, 127 (1998);
astro-ph/9711069
\bibitem{polardefect97}   Seljak U.,   Pen U.-L.,   Turok N.,   Phys.Rev.Lett. 79,  1615 (1997);  astro-ph/9704231
\bibitem{polar96}Kamionkowski, M., Kosowsky, A., and Stebbins, A., Phys. Rev. D 55, 7368 (1997);
astro-ph/9611125
\bibitem{polar98}Kamionkowski, M., and Kosowsky, A., 
Phys. Rev. D57, 685 (1998); astro-ph/9705219
\bibitem{pulsar94}Kaspi, V., Taylor, J., and Ryba, M., ApJ 428, 713 (1994)
\bibitem{sbbn95}Copi, C. J., Schramm, D. N., and Turner, M. S.,
Science 267, 192 (1995)
\bibitem{sbbn96}Hogan, C. J. in 
{\it Critical Dialogues in Cosmology}, 
 ed. N. Turok, (Princeton University, 1996);
astro-ph 9609138 
\bibitem{sarkar96} Sarkar, S., Rep. Prog. Phys. 59, 1493 (1996); hep-ph/9602260
\bibitem{sbbn97}Hata, N., Steigman, G., Bludman, S., Langacker, P.,
Phys. Rev. Lett. 55, 540 (1997)
\bibitem{sbbn98} Fiorentini, G., Lisi, E., Sarkar, S. and
Villante, F. L., to appear in Phys. Rev. D (1998); astro-ph/9803177
\bibitem{bdm94} Carr, B. J.   ARAA 32, 531 (1994)
\bibitem{inflation97}  Kolb, E. W.,
    in {\it Current
topic in Astrofundamental Physics,}
eds. N. Sanchez and A.
 Zichichi (World Scientific, 1997), p.162; astro-ph/9612138 
\bibitem{inflation97b} Lidsey, J. E.,  Liddle, A. R.,  Kolb, E. W.,
  Copeland E. J., Barreiro, T., and Abney, M.,
Rev. Mod. Phys. 69, 373 (1997)
\bibitem{peierls52}Peierls, R. E., Singwi, K. S. and Wroe, D.  PR 87,46, (1952)
\bibitem{nucleation83}  Hogan, C.~J.  
 Physics Letters 133B, 172,1983 
\bibitem{transition86}Hogan, 
C.~J., 
 M.N.R.A.S. 218,629 (1986)
\bibitem{transition90}Turner, M. S. and Wilcek, F., Phys. Rev. Lett. 65, 3080 (1990)
 \bibitem{transition93}  Ignatius J., Kajantie  K., Kurki-Suonio  H. and  Laine  M.
 Phys.Rev. D49,  3854 (1994); astro-ph/9309059
\bibitem{transition94}Kamionkowski,
 M., Kosowsky, A., and Turner, M. S., Phys.Rev. D 49, 2837  (1994)
\bibitem{transition96} Kurki-Suonio, H., and  Laine, M., Phys.Rev.Lett. 77,  3951 (1996); hep-ph/9607382
\bibitem{gold82}Vilenkin, A., Phys. Rev. Lett. 48, 59 (1982)
\bibitem{gold95}Hogan, C. J., Phys. Rev. Lett. 74, 3105, (1995);
astro-ph/9412054 
\bibitem{mtheory98}Schwarz, J. H., Phys. Rep. (1998); hep-th/9807135
\bibitem{strings81}Vilenkin, A., Phys. Lett. 107B, 47 (1981)
\bibitem{strings84}  Hogan, C.~J., and
Rees, M.~J.,  Nature  311, 109 , 1984   
\bibitem{strings85}Vilenkin, A., Phys. Rep. 121, 263 (1985)
\bibitem{strings94}Vilenkin, A. and Shellard, S., {\it Cosmic Strings and Other
Topological Defects}, Cambridge University Press (1994)
 \bibitem{strings97}    Battye, R. A.,
  Caldwell, R. R.,   Shellard, E. P. S.,
 to appear in {\it Topological Defects in Cosmology}, 
Ed. F.Melchiorri and M.Signore; astro-ph/9706013 
 \bibitem{strings98}     Avelino P.P.,  Shellard E.P.S.,  Wu, J.H.P., Allen  B.
      to appear in Phys. Rev. Lett. (Sept. 1998); astro-ph/9712008
\bibitem{defectCMB97}      Turok N.,   Pen U.-L.,   Seljak U., Phys.Rev. D58 (1998);
astro-ph/9706250
\bibitem{hybrid96}Martin, X. and  Vilenkin, A.,
Phys.Rev.Lett. 77, 2879 (1996); astro-ph/9606022
\end{references}
\end{document}